\documentclass[12pt]{article}

\usepackage{amsfonts}
\usepackage{amsmath}
\usepackage{amsthm}
\usepackage{cite}
\usepackage{epsfig}
\usepackage{latexsym}
\usepackage{paralist}
\usepackage{fancyhdr}
\usepackage{graphicx}
\numberwithin{equation}{section}
\usepackage[vcentermath]{youngtab}
\usepackage{young}
\usepackage{ytableau}
\usepackage{etex}
\usepackage{braket}
\usepackage{float}
\usepackage{autobreak}

\usepackage{pict2e}   
\usepackage{animate}  

\usepackage{multirow}
\usepackage{bigdelim}
\usepackage{fancybox}
\usepackage{cals}

\def\be{\begin{equation}}
\def\ee{\end{equation}}
\def\bea{\begin{eqnarray}}
\def\eea{\end{eqnarray}}

\renewcommand{\thefootnote}{\fnsymbol{footnote}}

\begin{document}

\hfuzz=100pt
\title{{\Large \bf{3d ``chiral'' Kutasov-Schwimmer duality} }}
\date{}
\author{ Keita Nii$^a$\footnote{nii@itp.unibe.ch}
}
\date{\today}

\maketitle

\thispagestyle{fancy}
\cfoot{}
\renewcommand{\headrulewidth}{0.0pt}

\vspace*{-1cm}
\begin{center}
$^{a}${{\it Albert Einstein Center for Fundamental Physics }}
\\{{\it Institute for Theoretical Physics
}}
\\ {{\it University of Bern}}  
\\{{\it  Sidlerstrasse 5, CH-3012 Bern, Switzerland}}

\end{center}

\begin{abstract}
We propose a ``chiral'' version of the Kutasov-Schwimmer duality in a 3d $\mathcal{N}=2$ $SU(N)$ gauge theory with $F$ fundamental matters, $\bar{F}$ anti-fundamental matters and an adjoint matter $X$ with a tree-level superpotential $W= \mathrm{tr} \, X^{k+1}$. The theory exhibits a rich structure of the baryonic and (dressed) Coulomb branch operators. At first sight, the duality seems bad due to the mismatch of the anti-baryonic branch in the moduli space of vacua. The duality well works by realizing that the anti-baryonic operators are identified with some of the dressed Coulomb branch coordinates under the proposed duality. This generalizes the $SU(N)$ ``chiral'' duality with (anti-)fundamental matters, which we previously proposed. 
\end{abstract}

\renewcommand{\thefootnote}{\arabic{footnote}}
\setcounter{footnote}{0}

\newpage
\tableofcontents 

\newpage

\section{Introduction}
Duality is a very powerful tool of studying the low-energy dynamics in strongly-coupled or non-perturbative gauge theories. In supersymmetric gauge theories with four supercharges, holomorphy highly constrains the dynamics \cite{Seiberg:1994bz} and the infrared duality is known as ``Seiberg duality'' \cite{Seiberg:1994pq}. Seiberg duality was first proposed in 4d $\mathcal{N}=1$ supersymmetric gauge theories with (anti-)fundamental flavors and then generalized to other theories with various tensor matters and theories in diverse dimensions. In this paper, we are interested in the Seiberg duality with adjoint matters, which is called Kutasov-Schwimmer duality \cite{Kutasov:1995ve, Kutasov:1995np, Kutasov:1995ss, Aharony:1995ne, Csaki:1998fm, Brodie:1996vx, Brodie:1996xm}.   

In three spacetime dimensions, Seiberg dualities have been well developed over the last two decades, including Chern-Simons dualities \cite{Giveon:2008zn}, $U(N)$ or $USp(2N)$ type dualities \cite{Aharony:1997gp} and so on. Recently, the method \cite{Aharony:2013dha, Aharony:2013kma} of deriving 3d dualities from 4d dualities via dimensional reduction was established and, by using this, the $SU(N)$ and $SO(N)$ type dualities were proposed.  
The 3d Seiberg dualities with tensor matters were studied in \cite{Kapustin:2011vz}. Especially, for adjoint matters, the dualities were investigated in \cite{Niarchos:2008jb, Niarchos:2009aa, Kim:2013cma, Park:2013wta, Hwang:2015wna, Hwang:2018uyj, Nii:2014jsa}. In most cases, the dualities were constructed for the ``vector-like'' theories with equal numbers of fundamental and anti-fundamental matters. The ``chiral'' theory with (anti-)fundamental matters was first studied in \cite{Benini:2011mf, Aharony:2014uya}. The chiral theory with adjoint matters was only considered for the $U(N)$ gauge group in \cite{Hwang:2015wna}. The low-energy dynamics of the $SU(N)$ chiral gauge theory only with (anti-)fundamental matters was studied in \cite{Aharony:2013dha, Intriligator:2013lca, Nii:2018bgf}.

In this paper, we develop the ``chiral'' version of the 3d $SU(N)$ Kutasov-Schwimmer duality. By elaborating on the Coulomb branch in the 3d $\mathcal{N}=2$ $SU(N)$ gauge theory with an adjoint matter, $F$ fundamental matters, $\bar{F}$ anti-fundamental matters and a tree-level superpotential $W_{ele}= \mathrm{tr} \, X^{k+1}$, we find that the dual description is the 3d  $\mathcal{N}=2$ $SU(kF-N)$ gauge theory with an adjoint matter, $F$ fundamental matters, $\bar{F}$ anti-fundamental matters, $k$ gauge singlet mesons and a tree-level superpotential $W_{mag}= \mathrm{tr} \, Y^{k+1} +\sum_{j=0}^{k-1} M_j  \tilde{q} Y^{k-1-j} q$. This duality is very similar to the conventional 4d $\mathcal{N}=1$ $SU(N)$ Kutasov-Schwimmer duality except for the fact that the 3d version must have ``chiral'' matter contents. Since the na\"ive dimensional reduction of 4d dualities does not work, one might consider that this 3d duality seems incorrect. For instance, this duality cannot have the mapping of the electric anti-baryons to the magnetic anti-baryons. As we will see in the main text, by carefully studying the Coulomb branch, we find that the anti-baryons are transformed to some of the Coulomb branch coordinates under the duality and thus the duality well works.

The rest of this paper is organized as follows. 
In Section 2, we will investigate the structure of the Coulomb branch in the 3d ``chiral'' $SU(N)$ gauge theories with (anti-)fundamental and adjoint matters. In order to describe the Coulomb moduli space, the ``dressed'' monopole operators are introduced. 
In Section 3, the 3d $\mathcal{N}=2$ $SU(N)$ ``chiral'' Kutasov-Schwimmer duality will be proposed.
In Section 4, we will show various concrete examples for small values of $N$. Some examples include the calculation of the superconformal indices and strengthen the duality proposal. 
In Section 5, we will summarize our findings and discuss future directions.

\section{Coulomb branch}
In this section, we investigate the structure of the Coulomb branch in 3d ``chiral'' $SU(N)$ gauge theories with adjoint matters in addition to (anti-)fundamental matters. While a lot of flat directions along the classical Coulomb branch are lifted by quantum corrections generated by monopole-instantons \cite{Aharony:1997bx, deBoer:1997kr}, some directions remain exactly massless. This situation is very different from the ``vector-like'' theories as noticed in \cite{Nii:2018bgf}. (See also \cite{Intriligator:2013lca, Aharony:2013dha} where the chiral theory was partially studied.) The theory of interest is a 3d $\mathcal{N}=2$ $SU(2N)$ gauge theory with an adjoint matter $X$, $F$ fundamental matters $Q$ and $\bar{F}$ anti-fundamental matters $\tilde{Q}$.  We here discuss only the $SU(2N)$ gauge group and assume $F>\bar{F}$. The generalization to the $SU(2N+1)$ case is straightforward. When the Coulomb branch denoted by $Y^{bare}_a$ obtains a non-zero expectation value, the gauge group and the matter fields are decomposed into
\begin{align}
SU(2N) & \rightarrow SU(N-a) \times SU(N-a) \times SU(2a) \times U(1)_1 \times U(1)_2  \\
{\tiny \yng(1)}& \rightarrow ({\tiny \yng(1)},1,1)_{1,a} +(1,{\tiny \yng(1)},1)_{-1,a} +(1,1,{\tiny \yng(1)})_{0,-(N-a)} \\
{\tiny \overline{\yng(1)}}&  \rightarrow ({\tiny \overline{\yng(1)}},1,1)_{-1,-a} +(1,{\tiny \overline{\yng(1)}},1)_{1,-a} +(1,1,{\tiny \overline{\yng(1)}})_{0,(N-a)} \\
\mathbf{adj.} &  \rightarrow (\mathbf{adj.},1,1)_{0,0}+(1,\mathbf{adj.} ,1)_{0,0}+(1,1,\mathbf{adj.} )_{0,0}+(1,1,1)_{0,0}+(1,1,1)_{0,0} \nonumber \\
&\qquad +({\tiny \yng(1)},{\tiny \overline{\yng(1)}},1)_{2,0} +({\tiny \overline{\yng(1)}},{\tiny \yng(1)},1)_{-2,0}+({\tiny \yng(1)},1,{\tiny \overline{\yng(1)}})_{1,N} + ({\tiny \overline{\yng(1)}},1,{\tiny \yng(1)})_{-1,-N} \nonumber \\
&\qquad \quad +(1,{\tiny \yng(1)},{\tiny \overline{\yng(1)}})_{-1,N} +(1,{\tiny \overline{\yng(1)}},{\tiny \yng(1)})_{1,-N},  \label{breaking1}
\end{align}
where $a=0,\cdots,N-1$ and the Coulomb branch $Y^{bare}_a$ corresponds to the first $U(1)_1$ subgroup. 
Since the Coulomb branch of the moduli space of vacua describes the flat direction of the scalar potential from the vector superfield, the Chern-Simons (CS) term $k_{eff}^{U(1)_1 U(1)_1}$ must be zero. Now, the breaking pattern is very symmetric and hence the CS term $k_{eff}^{U(1)_1 U(1)_1}$, which is for instance generated by the 1-loop diagrams of $({\tiny \yng(1)},1,1)_{1,a}$ and $(1,{\tiny \yng(1)},1)_{-1,a}$, is canceled out. Therefore, the ``chiral'' theory can have this type of the flat directions \cite{Nii:2018bgf}.

However, the whole story of the Coulomb branch is not so simple because other CS terms might be generated. In fact, the low-energy theory along the Coulomb branch obtains the mixed Chern-Simons terms between global and local $U(1)$ symmetries. We find that the mixed CS terms are generated as
\begin{align}
k_{eff}^{U(1)_1 U(1)_2} &=a(N-a) (F -\bar{F})\\
k_{eff}^{U(1)_1 Q} &= (N-a) \left[FQ +\bar{F} \bar{Q} +2(N+a)Q_{adj.}  \right],
\end{align}
where various $Q$'s denote the global charges of the matter (fermion) fields.
Since the matter content of the theory is ``chiral'' ($F > \bar{F}$), the bare Coulomb branch operator $Y_a^{bare}$ is not gauge invariant. The $U(1)_2$ charge of $Y_a^{bare}$ is proportional to $k_{eff}^{U(1)_1 U(1)_2}$ and it becomes $-a(N-a) (F -\bar{F})$ which is negative for $F > \bar{F}$. In order to construct the gauge invariant operator from it, we can multiply it by the massless component $(1,1,{\tiny \overline{\yng(1)}})_{0,(N-a)}$ of the anti-fundamental matters \cite{Csaki:2014cwa, Amariti:2015kha, Nii:2018bgf, Aharony:2015pla}. Since this component is charged under the remaining $SU(2a)$ gauge group, the bare Coulomb branch should be dressed by the generalized anti-baryon operators
\begin{gather}
\tilde{B}^{(n_0,\cdots, n_{k-1})} := \tilde{Q}^{n_0} (X \tilde{Q})^{n_1} \cdots (X^{k-1} \tilde{Q})^{n_{k-1}},~~~\sum n_\ell =2a \\
\tilde{Q} := (1,1,{\tiny \overline{\yng(1)}})_{0,(N-a)},~~~X:=(1,1,\mathbf{adj.} )_{0,0}.
\end{gather}
The dressed Coulomb branch operators are defined by
\begin{align}
Y^{dressed}_{a;n_0,\cdots, n_{k-1}} := Y_a^{bare} (\tilde{B}^{(n_0,\cdots, n_{k-1})})^{\frac{F-\bar{F}}{2}}.
\end{align}
Notice that $F-\bar{F}$ must be even due to the parity anomaly of the gauge symmetry. In order to construct the $SU(2a)$ anti-baryons, $a$ must satisfy
\begin{align}
2a \le k \bar{F}. \label{acon1}
\end{align}

The further constraint on $a$ comes from the stability of the supersymmetric vacuum of the low-energy $SU(N-a)$ theory with an adjoint matter. The $SU(N-a)$ gauge theory obtains a Chern-Simons level $\frac{F-\bar{F}}{2}$ and includes no (anti-)fundamental matters which are all massive and integrated out from the low-energy spectrum. Since we are looking for the flat directions of the theory, the low-energy $SU(N-a)$ theory must have a stable and supersymmetric vacuum. This requires\footnote{This constraint was first derived from the s-rule of the corresponding D-brane configuration \cite{Hanany:1996ie}.} \cite{Niarchos:2008jb, Niarchos:2009aa} 
\begin{align}
k \frac{F -\bar{F}}{2} \ge N-a. \label{acon2}
\end{align}
The (dressed) Coulomb branch $Y^{dressed}_{a;n_0,\cdots, n_{k-1}} $ can exist only when $a$ satisfies \eqref{acon1} and \eqref{acon2}. 

The quantum picture of the dressed Coulomb branch needs a more careful analysis. Since, along the gauge symmetry breaking \eqref{breaking1}, the adjoint matter supplies several massless (and neutral) components, the Coulomb branch can be also dressed by these massless scalar fields. We should generally take into account those Coulomb branch operators. In addition, these operators dressed by the massless components of $X$ are truncated classically and non-perturbatively. Here, we will not discuss the full quantum picture of the (dressed) Coulomb branch but the analysis in this section helps us understand the ``chiral'' version of the 3d Kutasov-Schwimmer duality.

\section{3d $SU(N)$ ``chiral'' Kutasov-Schwimmer duality}
Here, we propose the ``chiral'' version of the 3d $SU(N)$ Kutasov-Schwimmer duality. 
The electric description is the 3d $\mathcal{N}=2$ $SU(N)$ gauge theory with an adjoint matter $X$, $F$ fundamental matters $Q$ and $\bar{F}$ anti-fundamental matters $\tilde{Q}$. We assume $F>\bar{F}$ without loss of generality. The electric theory has a tree-level superpotential
\begin{align}
W_{ele} = \mathrm{tr} \, X^{k+1},
\end{align}
which truncates and simplifies the chiral ring structure. The F-flatness condition imposes $X^k =\frac{1}{N}  \mathrm{tr} \, X^k $. The quantum numbers of the electric matter content are summarized in Table \ref{SU(N)KSelectric}. The charge assignment of the infrared R-symmetry is ambiguous and we listed the generic r-charges in Table \ref{SU(N)KSelectric}. The Higgs branch is identical to the 4d one although the meson operators are modified to non-square matrices
\begin{align}
M_{j}:=QX^j \tilde{Q},~~~~j=0,\cdots,k-1.
\end{align}
The generalized baryonic operators are defined by
\begin{align}
B^{(n_0,n_1,\cdots,n_{k-1})} &:=Q_{(0)}^{n_0} Q_{(1)}^{n_1} \cdots Q_{(k-1)}^{n_{k-1}},~~~~\sum_{\ell=0}^{k-1} n_\ell =N\\
Q_{(\ell)} &:= X^\ell Q,
\end{align}
where the gauge indices are contracted by an epsilon tensor. The power of $X$ is truncated at $O(X^{k-1})$ due to the F-flatness condition. The anti-baryonic operators are also defined in the same manner. We also have the Higgs branch coordinates where only the adjoint scalar obtains a non-zero expectation value
\begin{align}
T_i:=\mathrm{tr} \,X ^i,~~~~(i=2,\cdots,k).
\end{align}
These operators (including the Coulomb branch operators) are not linearly independent and there are some constraints relating them since these operators are composite operators of the elementary fields. The constraints appear both classically and quantum-mechanically. In order to rigorously prove the duality and test the matching of the chiral ring, we have to consider the mapping of these operators including the quantum constraints. In this paper, however, we will mostly focus on the na\"ive matching of these gauge invariant operators and this will serve as a first non-trivial check of the duality. For an additional support, we will investigate the superconformal indices of the duality pair in the subsequent section.

\begin{table}[H]\caption{3d $\mathcal{N}=2$ $SU(N)$ with $adj. + F\, {\tiny \protect\yng(1)}+ \bar{F} \,{\tiny \overline{\protect\yng(1)}}$, $W_{ele} = \mathrm{tr} \, X^{k+1}$} 
\begin{center}
\scalebox{0.9}{
  \begin{tabular}{|c||c||c|c|c|c|c| } \hline
  &$SU(N)$&$SU(F)$&$SU(\bar{F})$&$U(1)$&$U(1)$&$U(1)_R$  \\ \hline
 $X$ &$\mathbf{adj.}$&1&1&0&0&$\frac{2}{k+1}$ \\
 $Q$ & ${\tiny \yng(1)}$ &${\tiny \yng(1)}$&1&1&0&$r$ \\
$\tilde{Q}$  &${\tiny \overline{\yng(1)}}$&1&${\tiny \yng(1)}$&$0$&1&$\bar{r}$ \\  \hline
$M_{j=0,\cdots,k-1}:=QX^j \tilde{Q}$&1&${\tiny \yng(1)}$&${\tiny \yng(1)}$&1&1&$r+\bar{r} +\frac{2j}{k+1}$  \\
$T_i:=\mathrm{tr} \,X ^i$&1&1&1&0&0&$\frac{2i}{k+1}$  \\ \hline
  \end{tabular}}
  \end{center}\label{SU(N)KSelectric}
\end{table}

Let us investigate the ``chiral'' Kutasov-Scwimmer dual of the electric theory in Table \ref{SU(N)KSelectric}. We propose that the dual description is given by the 3d $\mathcal{N}=2$ $SU(kF-N)$ gauge theory with an adjoint matter $Y$, $F$ (dual) fundamental matters $q$, $\bar{F}$ (dual) anti-fundamental matters $\tilde{q}$ and the meson gauge singlets $M_j$ $(j=0,\cdots,k-1)$. The magnetic theory has a tree-level superpotential
\begin{align}
W_{mag} =\mathrm{tr} \, Y^{k+1} +\sum_{j=0}^{k-1} M_j  \tilde{q} Y^{k-1-j} q,
\end{align}
where we omitted the dimensionful couplings for simplicity. 
The quantum numbers of the magnetic matter content are summarized in Table \ref{SU(N)KSmagnetic}. The charge assignment is completely fixed by the above superpotential and by requiring the matching of the generalized baryonic operators as follows. 
\begin{align}
q_{(\ell)} &:=Y^\ell q \\
b^{(m_0,m_1,\cdots,m_{k-1})} &:= q_{(0)}^{m_0}q_{(1)}^{m_1} \cdots q_{(k-1)}^{m_{k-1}},~~~~\sum_{\ell=0}^{k-1} m_\ell =kF-N \\
B^{(n_0,n_1,\cdots,n_{k-1})}& \sim b^{(m_0,m_1,\cdots,m_{k-1})},~~~m_\ell =F-n_{k-1-\ell}
\end{align}
As we will see below, the remaining moduli coordinates which include the anti-baryonic operators and the dressed Coulomb branch can have a correct transformation law under the duality. This would be a first non-trivial check of this duality because we only required consistency of the charge assignment in the superpotential and the baryon matching.

\begin{table}[H]\caption{$SU(kF-N)$ magnetic dual description} 
\begin{center}
\scalebox{0.9}{
  \begin{tabular}{|c||c||c|c|c|c|c| } \hline
  &$SU(kF-N)$&$SU(F)$&$SU(\bar{F})$&$U(1)$&$U(1)$&$U(1)_R$  \\ \hline
 $Y$ &$\mathbf{adj.}$&1&1&0&0&$\frac{2}{k+1}$ \\
 $q$ & ${\tiny \yng(1)}$ &${\tiny \overline{\yng(1)}}$&1&$\frac{N}{kF-N}$&0&$\frac{N}{kF-N}r -\frac{k-1}{k+1}\frac{kF-2N}{kF-N}$ \\
$\tilde{q}$  &${\tiny \overline{\yng(1)}}$&1&${\tiny \overline{\yng(1)}}$&$-\frac{kF}{kF-N}$&$-1$&$-\frac{kF}{kF-N}r-\bar{r}+2-\frac{k-1}{k+1}\frac{kF}{kF-N}$ \\ 
$M_j$&1&${\tiny \yng(1)}$&${\tiny \yng(1)}$&1&1&$r+\bar{r} +\frac{2j}{k+1}$  \\  \hline
$T_a\sim \mathrm{tr} \,Y ^a$&1&1&1&0&0&$\frac{2a}{k+1}$  \\ \hline
  \end{tabular}}
  \end{center}\label{SU(N)KSmagnetic}
\end{table}

Let us study the matching of the gauge invariant operators under the duality. Especially, we will focus on the transformation of the baryon and (dressed) Coulomb branch operators. In order to make the discussion explicit, we consider the electric $SU(2N)$ gauge theory with $F$ fundamentals and $\bar{F}$ anti-fundamentals. In addition, we take $k=2n$. The dual gauge group becomes $SU(2nF -2N)$ and we can apply the analysis in the previous section to the electric and magnetic sides. Depending on the values of $N, F, \bar{F}$ and $n$, the content of the possible baryon and Coulomb branch operators drastically change. Furthermore, the chiral ring structure will be modified classically and quantum-mechanically. Here, we only list the generic gauge invariant operators and hence some operators are not available for a particular combination of $(N, F, \bar{F}, n)$. 

On the electric side, we can define the (anti-)baryonic and Coulomb branch operators 
\begin{align}
B^{(n_0,\cdots,n_{2n-1})} &:=Q_{(0)}^{n_0} Q_{(1)}^{n_1} \cdots Q_{(2n-1)}^{n_{2n-1}} \\
\bar{B}^{(\bar{n}_0, \cdots,\bar{n}_{2n-1})} &:=\tilde{Q}^{\bar{n}_0}_{(0)} \tilde{Q}_{(1)}^{ \bar{n}_1} \cdots \tilde{Q}_{(2n-1)}^{\bar{n}_{2n-1}}   \\
Y_{a_e; p_0,\cdots,p_{2n-1}}^{dressed} &:= Y_{a_e}^{bare}(\tilde{B}^{(p_0,\cdots, p_{2n-1})})^{\frac{F-\bar{F}}{2}},
\end{align}
where $\sum_j n_j =\sum_j \bar{n}_j =2N$ and $\sum_j p_j = 2a_e$.
For $a_e=N-n(F-\bar{F})$, the bare Coulomb branch operator with a more minimal magnetic charge
 can be dressed and made gauge invariant by a generalized anti-baryon operator as follows.
\begin{align}
 Y^{\frac{2}{F-\bar{F}}, dressed}_{a_e=N-n(F-\bar{F}); \left\{\bar{p}_j \right\} } := \left( Y^{bare}_{a_e=N-n(F-\bar{F})} \right)^{\frac{2}{F-\bar{F}}} \tilde{Q}^{\bar{p_0}} (X \tilde{Q})^{\bar{p_1}} \cdots (X^{2n-1} \tilde{Q})^{\bar{p}_{2n-1}}. \label{minimalmonopole}
\end{align}
In this case, $Y_{a_e=N-n(F-\bar{F}); p_0,\cdots,p_{2n-1}}^{dressed}$ should be regarded as a composite operator of the minimal one. In Table \ref{BaryonMonopole}, the $U(1)$ charges of these fields are summarized.

On the magnetic side, one can similarly define the dual (anti-)baryons and the magnetic Coulomb branch operators
\begin{align}
b^{(m_0,\cdots,m_{2n-1})} &:= q_{(0)}^{m_0} q_{(1)}^{m_1} \cdots q_{(2n-1)}^{m_{2n-1}}  \\
\bar{b}^{(\bar{m}_0, \cdots,\bar{m}_{2n-1})} &:= \tilde{q}^{\bar{m}_0}_{(0)} \tilde{q}_{(1)}^{ \bar{m}_1}\cdots \tilde{q}_{(2n-1)}^{\bar{m}_{2n-1}}   \\
\tilde{Y}_{a_m; q_0,\cdots,q_{2n-1}}^{dressed} &:=   \tilde{Y}_{a_m}^{bare}(\tilde{b}^{(q_0,\cdots, q_{2n-1})})^{\frac{F-\bar{F}}{2}}.
\end{align}
For $a_m= n\bar{F}-N$, one can construct the more minimal Coulomb branch operator as \eqref{minimalmonopole}, which is denoted by $\tilde{Y}^{\frac{2}{F-\bar{F}},dressed}_{a_m=n \bar{F}-N; \left\{ \bar{q}_j \right\}}$. The $U(1)$ charges of these operators are also summarized in Table \ref{BaryonMonopole}. The set of the non-negative integers $\left\{n_j, \bar{n_j}, m_j, \bar{m}_j, p_j, \bar{p}_j, q_j, \bar{q}_j | j=0,\cdots,2n-1 \right\}$ satisfies the following constraints
\begin{gather*}
\sum_j n_j =\sum_j \bar{n}_j =2N,~~~\sum_j m_j =\sum_j \bar{m}_j =2nF -2N  \\
\sum_j p_j = 2a_e,~~~\sum_j \bar{p}_j=2N-2n(F-\bar{F}),~~~\sum_j q_j =2a_m,~~~\sum_j \bar{q}_j=2n \bar{F} -2N.
\end{gather*}
From Table \ref{BaryonMonopole}, we can easily find that the matching of the baryon and (dressed) Coulomb branch operators can be achieved by substituting the following relations  
\begin{gather*}
a_m = n\bar{F} -a_e,~~~m_j= F-n_{2n-1-j},~~~\bar{m}_j=\bar{F} -\bar{p}_{2n-1-j} \\
q_j =\bar{F} -p_{2n-1-j},~~~\bar{q}_j= \bar{F} -\bar{n}_{2n-1-j}.
\end{gather*}
The mapping of the gauge invariant operators in Table \ref{BaryonMonopole} becomes
\begin{gather*}
B^{(n_0,\cdots,n_{2n-1})} \leftrightarrow b^{(m_0,\cdots,m_{2n-1})},~~~\bar{B}^{(\bar{n}_0, \cdots,\bar{n}_{2n-1})} \leftrightarrow \tilde{Y}^{\frac{2}{F-\bar{F}},dressed}_{a_m=n \bar{F}-N; \left\{ \bar{q}_j \right\}} \\
Y_{a_e; p_0,\cdots,p_{2n-1}}^{dressed} \leftrightarrow \tilde{Y}^{dressed}_{a_m; q_0,\cdots,q_{2n-1} },~~~Y^{\frac{2}{F-\bar{F}}, dressed}_{a_e=N-n(F-\bar{F}); \left\{\bar{p}_j \right\} } \leftrightarrow \bar{b}^{(\bar{m}_0,\cdots, \bar{m}_{2n-1})}.
\end{gather*}

\begin{table}[H]\caption{$U(1)$ charges of baryons and Coulomb branch operators $(r=\bar{r}=0)$} 
\begin{center}
\scalebox{0.65}{
  \begin{tabular}{|c||c|c|c| } \hline
  &$U(1)$&$U(1)$&$U(1)_R$  \\ \hline
  $B^{(n_0,\cdots,n_{2n-1})}$&$2N$&0&$\frac{2}{2n+1}\sum j n_j$  \\
  $\bar{B}^{(\bar{n}_0, \cdots,\bar{n}_{2n-1})}$&0&$2N$&$\frac{2}{2n+1}\sum j \bar{n}_j$  \\
 $Y_{a_e; p_0,\cdots,p_{2n-1}}^{dressed}$ &$-FN+Fa_e$&$-\bar{F}N+Fa_e$&$(N-a_e) \left(F+\bar{F} -\frac{4}{2n+1}(N+a_e)\right) +\frac{F-\bar{F}}{2n+1}\sum j p_j $  \\ 
 $Y^{\frac{2}{F-\bar{F}}, dressed}_{a_e=N-n(F-\bar{F}); \left\{\bar{p}_j \right\} }$&$-2n F$&$-2nF+2N$&$4nF-4N+\frac{2n-1}{2n+1}(2n(F-\bar{F} )-4N)+\frac{2}{2n+1}\sum j \bar{p}_j$ \\ \hline 
  $b^{(m_0,\cdots,m_{2n-1})}$&$2N$&0&\footnotesize $-\frac{2n-1}{2n+1}(2nF-4N) +\frac{2}{2n+1}\sum j m_j = \frac{2}{2n+1} \sum j n_j $  \\
  $\bar{b}^{(\bar{m}_0,\cdots, \bar{m}_{2n-1})}$&$-2nF$&$-2nF+2N$&\footnotesize $4nF-4N -\frac{2n-1}{2n+1}2nF +\frac{2}{2n+1} \sum j \bar{m}_j $  \\
 $\tilde{Y}^{dressed}_{a_m; q_0,\cdots,q_{2n-1} }$ &\small $-FN+(n \bar{F}-a_m)F$&$-\bar{F}N +(n \bar{F}-a_m)F$&\footnotesize $(N-(n\bar{F}-a_m)) \left(F+\bar{F} -\frac{4}{2n+1}(N+(n\bar{F}-a_m))\right) +\frac{F-\bar{F}}{2n+1}\sum j (\bar{F}-q_{2n-1-j})  $ \\
  $\tilde{Y}^{\frac{2}{F-\bar{F}},dressed}_{a_m=n \bar{F}-N; \left\{ \bar{q}_j \right\}}$&0&$2N$& $\frac{2}{2n+1} \sum j (\bar{F}-\bar{q}_{2n-1-j}) $     \\ \hline
  \end{tabular}}
  \end{center}\label{BaryonMonopole}
\end{table}

We can test various deformations of the duality proposed above. 
By gauging the $U(1)_B$ symmetry which is a linear combination of the two $U(1)$ global symmetries, we obtain the ``chiral'' $U(N)$ duality which was studied in \cite{Hwang:2015wna}. Since the un-gauging procedure is generally non-trivial, this can be regarded as the derivation of the $U(N)$ duality from the $SU(N)$ duality. For $k=1$, the adjoint matter is massive and integrated out.  As a result, the duality goes back to the ``chiral'' $SU(N)$ duality studied in \cite{Nii:2018bgf}. By introducing the complex mass to the (anti-)fundamental matter, say, $W_{ele}=m M_{0}^{F,\bar{F}}$, the electric side flows to the $SU(N)$ gauge theory with $(F-1, \bar{F}-1)$ (anti-)fundamentals. On the magnetic side, the gauge group is higgsed to $SU(kF-k-N )$ and one (dual) flavor is eaten. In this way, the duality is correctly preserved under the complex mass deformation with reduction of $F$ and $\bar{F}$.  

Finally, we consider the weak deformation of the tree-level superpotential by powers of the adjoint chiral superfield
\begin{align}
W= \sum_{j=2}^{k} g_j \, \mathrm{tr} \, X^{j+1}+\lambda \, \mathrm{tr} \,X,
\end{align}
where the traceless condition is imposed by the Lagrange multiplier $\lambda$.
The same deformation is also turned on the dual side by using $Y$. For generic values of $g_j$, the adjoint matter obtains the $k$ distinct eigenvalues and the gauge group is higgsed into 
\begin{align}
SU(N) \rightarrow SU(N_1) \times \cdots SU(N_{k}) \times U(1)^{k-1},~~~~\sum_{i=1}^{k}N_i =N. 
\end{align}
The adjoint matters are massive along this breaking and integrated out. The low-energy theory for each $SU(N_i)$ becomes the 3d $\mathcal{N}=2$ $SU(N_i)$ gauge theory with $F$ fundamentals and $\bar{F}$ anti-fundamentals. 
The dual of the ``chiral'' $SU(N_i)$ SQCD was proposed in \cite{Nii:2018bgf} and given by the $SU(F-N_i)$ gauge theory. On the magnetic side, the similar breaking takes place and we obtain the $SU(\tilde{N}_i)$ gauge theory with $\sum_{i=1}^{k} \tilde{N}_i=kF-N$. By identifying $\tilde{N}_i =F- N_i$, the ``chiral'' Kutasov-Schwimmer duality correctly reduces to the ``chiral'' Seiberg duality \cite{Nii:2018bgf} for each $SU(N_i)$ factor.

\section{Examples and Superconformal Indices}
In this section, we will show various concrete examples of the 3d $SU(N)$ ``chiral'' Kutasov-Schwimmer duality. In the previous section, we focused on the generic structure of the proposed duality, paying attention to the (na\"ive) matching of the moduli operators. Since the content of the gauge invariant operators changes classically and quantum-mechanically, it would be valuable to explicitly investigate several examples. For some cases, we will compute the superconformal indices \cite{Bhattacharya:2008bja, Kim:2009wb, Imamura:2011su, Kapustin:2011jm} and verify the validity of the proposed duality.

\subsection{Completely ``chiral'' limit: $\bar{F}=0$}
In this subsection, we consider the chiral limit of the proposed duality, where we will take $\bar{F}=0$. 
The theory is completely ``chiral''. For the $SU(2N+1)$ gauge groups, the theory cannot have the Coulomb flat directions since the dressed operators are not available. For the $SU(2N)$ gauge groups, the Coulomb branch is available only for $a=0$. Due to the parity anomaly constraint of the gauge symmetry, the number of the fundamental matters must be even.

\subsubsection{$SU(2N)$ examples}
We start with the completely chiral case for the $SU(2N)$ gauge groups. The electric theory is the 3d $\mathcal{N}=2$ $SU(2N)$ gauge theory with $2F$ fundamentals and an adjoint matter. The superpotential is $W_{ele} = \mathrm{tr} \, X^{k+1}$. The similar theory without an adjoint matter was studied in \cite{Aharony:2013dha} and the corresponding Coulomb branch was studied. The Coulomb branch, denoted by  $Y^{bare}_{a=0}$, corresponds to the breaking 
\begin{align}
SU(2N) \rightarrow SU(N) \times SU(N)' \times U(1),
\end{align}
where $Y^{bare}_{a=0}$ is a dualized chiral superfield of the $U(1)$ subgroup. Since the low-energy theory only includes a single $U(1)$ subgroup, the bare Coulomb branch operator is gauge invariant. Therefore, the lowest Coulomb branch operator is $Y^{bare}_{a=0}$. In addition, since the adjoint scalar reduces to the massless adjoint fields in $SU(N) \times SU(N)' \times U(1)$, the dressed Coulomb branch operators are also possible:
\begin{align}
Y_{a=0,X_{U(1)}}^{dressed} & :=Y_{a=0}^{bare} X_{U(1)} \\
Y_{a=0,X_{SU(N)}}^{dressed} &:=Y_{a=0}^{bare} \mathrm{tr}\,X^2_{SU(N)} \\
Y_{a=0,X_{SU(N)'}}^{dressed} &:=Y_{a=0}^{bare} \mathrm{tr}\,X^2_{SU(N)'},
\end{align}
where $Y_{a=0,X_{U(1)}}^{dressed}$ is dressed by a single $U(1)$ adjoint scalar which is a gauge singlet. For $Y_{a=0,X_{SU(N)}}^{dressed}$ and $Y_{a=0,X_{SU(N)'}}^{dressed} $, the bare operator is dressed by the two $SU(N)$ adjoint scalars. For $N>2$, the more highly dressed operators are generally possible although they might be classically related to each other and quantum-mechanically truncated due to non-perturbative effects. In this paper, we will not consider the independent set of the moduli coordinates but we expect that quantum effects will truncate these dressed operators and hence the duality well works even if the rank of the gauge group differs from the magnetic one. The quantum analysis of the chiral ring would be left as a future work. Table \ref{SU(2N)chiralelectric} summarizes the quantum numbers of the electric matter content and the Coulomb branch operators. 

\begin{table}[H]\caption{3d $\mathcal{N}=2$ $SU(2N)$ with $\mathbf{adj.} + 2F\, {\tiny \protect\yng(1)}$ and $W_{ele} = \mathrm{tr}\, X^{k+1}$} 
\begin{center}
\scalebox{1}{
  \begin{tabular}{|c||c||c|c|c| } \hline
  &$SU(2N)$&$SU(2F)$&$U(1)$&$U(1)_R$  \\ \hline
 $X$ &$\mathbf{adj.}$&1&0&$\frac{2}{k+1}$ \\
 $Q$ & ${\tiny \yng(1)}$ &${\tiny \yng(1)}$&1&$r$ \\  \hline 
 $Y^{bare}_{a=0}$&1&1&$-2FN$&$-2FNr +2FN-\frac{4N^2}{k+1}$  \\  
$Y_{a=0,X_{U(1)}}^{dressed} $ &1&1&$-2FN$&$-2FNr +2FN -\frac{4N^2 -2}{k+1}$  \\
$Y_{a=0,X_{SU(N)}}^{dressed}$ &1&1&$-2FN$&$-2FNr +2FN -\frac{4N^2 -4}{k+1}$   \\
$Y_{a=0,X_{SU(N)'}}^{dressed}$ &1&1&$-2FN$&$-2FNr +2FN -\frac{4N^2 -4}{k+1}$     \\ \hline
   \end{tabular}}
  \end{center}\label{SU(2N)chiralelectric}
\end{table}

On the other hand, the magnetic description becomes the 3d $\mathcal{N}=2$ $SU(2kF-2N)$ gauge theory with $2F$ (dual) fundamental matters and an adjoint matter. The magnetic superpotential only includes $W_{mag} = \mathrm{tr} \, Y^{k+1}$ since there is no gauge singlet elementary field. The quantum numbers of the magnetic fields are summarized in Table \ref{SU(2N)chiralmagnetic}. The analysis of the magnetic Coulomb branch can be performed in the same manner as the electric side. The bare Coulomb branch $\tilde{Y}^{bare}_{a=0}$ corresponds to the gauge symmetry breaking $SU(2kF-2N) \rightarrow SU(kF-N) \times SU(kF-N)' \times U(1)$ and is dressed by the adjoint scalars. 
From the $U(1)$ charges of the Coulomb branch operators, we can easily see that the magnetic Coulomb branch is mapped to the electric one. The baryon operators are also mapped to the (dual) baryons.

\begin{table}[H]\caption{$SU(2kF-2N)$ magnetic dual of Table \ref{SU(2N)chiralelectric}} 
\begin{center}
\scalebox{0.83}{
  \begin{tabular}{|c||c||c|c|c| } \hline
  &$SU(2kF-2N)$&$SU(2F)$&$U(1)$&$U(1)_R$  \\ \hline
 $Y$ &$\mathbf{adj.}$&1&0&$\frac{2}{k+1}$ \\
 $q$ & ${\tiny \yng(1)}$ &${\tiny \overline{\yng(1)}}$&$\frac{N}{kF-N}$&$\frac{N}{kF-N}r-\frac{k-1}{k+1}\frac{kF-2N}{kF-N} $ \\  \hline 
 $\tilde{Y}^{bare}_{a=0}$&1&1&$-2FN$&$-2FNr +2FN-\frac{4N^2}{k+1}$  \\  
$\tilde{Y}^{dressed}_{a=0,Y_{U(1)}}:=\tilde{Y}^{bare}_{a=0} Y_{U(1)} $&1&1&$-2FN$&$-2FNr +2FN-\frac{4N^2-2}{k+1}$  \\
$\tilde{Y}^{dressed}_{a=0,Y_{SU(kF-N)}}:=\tilde{Y}^{bare}_{a=0} \mathrm{tr}\, Y_{SU(kF-N)}^2 $&1&1&$-2FN$&$-2FNr +2FN-\frac{4N^2-4}{k+1}$  \\
$\tilde{Y}^{dressed}_{a=0,Y_{SU(kF-N)'}}:=\tilde{Y}^{bare}_{a=0} \mathrm{tr}\, Y_{SU(kF-N)'}^2 $&1&1&$-2FN$&$-2FNr +2FN-\frac{4N^2-4}{k+1}$  \\  \hline
   \end{tabular}}
  \end{center}\label{SU(2N)chiralmagnetic}
\end{table}

\subsubsection{$SU(2N+1)$ examples}
As an example of the $SU(2N+1)$ duality, let us consider the 3d $\mathcal{N}=2$ $SU(3)$ gauge theory with an adjoint matter and two fundamental matters. By setting $k=2$, the theory includes a tree-level superpotential 
\begin{align}
W_{ele}= \mathrm{tr} \, X^3. \label{Wk2}
\end{align}
Table \ref{SU(3)confinement} summarizes the quantum numbers of the elementary fields.
Since the gauge group is $SU(3)$ and there is no anti-fundamental matter, the dressed Coulomb branch operator is not available. The Higgs branch is also simplified as noticed in the 4d examples \cite{Kutasov:1995ss, Csaki:1998fm}. Especially, the moduli operator constructed only from $X$ is not available \cite{Csaki:1998fm} and the baryonic operator is defined only for $B^{(2,1)}:=Q^2 (XQ)$. Therefore, the low-energy dynamics is expected to be dual to a single free baryon $B^{(2,1)}$ which has two components. This is consistent with our duality proposal since the dual gauge group is now vanishing and the baryon $B^{(2,1)}$ is identified with the dual quark $q$.

\begin{table}[H]\caption{3d $\mathcal{N}=2$ $SU(3)$ with $\mathbf{adj.} + 2\, {\tiny \protect\yng(1)}$, $W_{ele}= \mathrm{tr} \, X^3$} 
\begin{center}
\scalebox{1}{
  \begin{tabular}{|c||c||c|c|c| } \hline
  &$SU(3)$&$SU(2)$&$U(1)$&$U(1)_R$  \\ \hline
 $X$ &$\mathbf{adj.}$&1&0&$\frac{2}{3}$ \\
 $Q$ & ${\tiny \yng(1)}$ &${\tiny \yng(1)}$&1&$r$ \\  \hline 
 $B^{(2,1)}:=Q^2 (XQ)$&1&${\tiny \yng(1)}$&3&$3r +\frac{2}{3}$  \\ \hline
  \end{tabular}}
  \end{center}\label{SU(3)confinement}
\end{table}

As a check of the duality, we can compare the superconformal indices of the electric and magnetic theories. We will focus on the specific r-charge, let's say $r=\frac{1}{6}$, and introduce the fugacity only for the $U(1)$ global symmetry, which is denoted by $t$. By employing the localization results \cite{Bhattacharya:2008bja, Kim:2009wb, Imamura:2011su, Kapustin:2011jm}, we compute the SCI of the $SU(3)$ electric theory as follows
\small
\begin{align}
I &=1-\frac{2 x^{5/6}}{t^3}+2 t^3 x^{7/6}+\frac{x^{5/3}}{t^6}-4 x^2+3 t^6 x^{7/3}-4 t^3 x^{19/6}+4 t^9 x^{7/2}+\frac{4 x^{11/3}}{t^6}-5 x^4+\cdots \nonumber \\
&=  \left.\left(   \frac{\left( t^{-3} x^{2- \left(3r +\frac{2}{3} \right)};x^2\right)_{\infty }}{\left(t^3 x^{\left(3r +\frac{2}{3} \right)};x^2\right)_{\infty }}  \right)^2   \right|_{r=\frac{1}{6}},
\end{align}
\normalsize where $(a;q)_{\infty}$ is a q-Pochhammer symbol. On the second line, the series is represented as the index of the single baryon operator $B^{(2,1)}$ and we find that the theory is indeed s-confining. We can see that the boson contribution comes only from  $B^{(2,1)}$ and that the operators like $\mathrm{tr} \, X^2$ and $Q(XQ)^2$ are removed from the chiral ring.

\subsection{$SU(2)$ duality}
 We here investigate the $SU(2)$ duality example whose Coulomb branch is simple enough to understand the matching of the baryonic and Coulomb branch operators under the duality transformation. The electric description is the 3d $\mathcal{N}=2$ $SU(2)$ gauge theory with $F+\bar{F}$ fundamentals and an adjoint matter. In order to describe the ``chiral'' duality, the fundamental matters (doublets) are decomposed into $F$ fundamental and $\bar{F}$ anti-fundamental matters. Since the theory must not have the parity anomaly for the gauge symmetry, $F+\bar{F}$ should be even. For simplicity, we set $k=2n$ but one can easily generalize our analysis to the odd $k$ case. Since we are considering the $SU(2)$ gauge group, the tree-level superpotential is vanishing. Table \ref{SU(2)KSelectric} summarizes the quantum numbers of the electric matter fields and the moduli operators. The Higgs branch is described by the following operators
\begin{gather*}
M_{j}:=QX^j \tilde{Q}~~(j=0,1),~~~T_2:=\mathrm{tr} \,X ^2,~~~B^{(2,0)} :=Q^2\\
B^{(1,1)} :=QXQ,~~~\bar{B}^{(2,0)}:= \tilde{Q}^2,~~~\bar{B}^{(1,1)}:= \tilde{Q} X \tilde{Q}.
\end{gather*}
Since the gauge group is $SU(2)$, the Higgs branch operators are truncated at $O(X^2)$. The vev of the Coulomb branch induces  the gauge symmetry breaking
\begin{align}
SU(2) & \rightarrow U(1) \\ 
\mathbf{2} & \rightarrow \mathbf{1}_{+1} + \mathbf{1}_{-1}  \\
\mathbf{3} & \rightarrow \mathbf{1}_{+2} +\mathbf{1}_{0} +\mathbf{1}_{-2}.
\end{align}
Since the adjoint representation reduces to the massless singlet $\mathbf{1}_{0}$ along the Coulomb branch, we can dress the bare monopole (Coulomb branch) operator by $\mathbf{1}_{0}$. The resulting dressed operator is defined by
\begin{align}
 Y_{SU(2),X} := Y_{SU(2)}^{bare} \mathbf{1}_{0} \sim Y_{SU(2)}^{bare} X. \label{SU2u1dressed}
\end{align}
In addition to this, the bare operator $Y_{SU(2)} := Y_{SU(2)}^{bare} $ is also gauge invariant and becomes a chiral ring element on the Coulomb branch. Therefore, the Coulomb branch is two-dimensional. Notice that the fundamental matters are massive along the Coulomb branch and hence there is no Coulomb branch dressed by the fundamental matters.

\begin{table}[H]\caption{3d $\mathcal{N}=2$ $SU(2)$ with $\mathbf{adj.} + F\, {\tiny \protect\yng(1)}+ \bar{F} \,{\tiny \overline{\protect\yng(1)}}$} 
\begin{center}
\scalebox{0.9}{
  \begin{tabular}{|c||c||c|c|c|c|c| } \hline
  &$SU(2)$&$SU(F)$&$SU(\bar{F})$&$U(1)$&$U(1)$&$U(1)_R$  \\ \hline
 $X$ &$\mathbf{adj.}$&1&1&0&0&$\frac{2}{2n+1}$ \\
 $Q$ & ${\tiny \yng(1)}$ &${\tiny \yng(1)}$&1&1&0&$r$ \\
$\tilde{Q}$  &${\tiny \overline{\yng(1)}}$&1&${\tiny \yng(1)}$&$0$&1&$\bar{r}$ \\  \hline
$M_{j=0,1}:=QX^j \tilde{Q}$&1&${\tiny \yng(1)}$&${\tiny \yng(1)}$&1&1&$r+\bar{r} +\frac{2j}{2n+1}$  \\
$T_2:=\mathrm{tr} \,X ^2$&1&1&1&0&0&$\frac{4}{2n+1}$  \\ 
$B^{(2,0)} :=Q^2$&1&${\tiny \yng(1,1)}$&1&2&0&$2r$  \\
$B^{(1,1)} :=QXQ$&1&${\tiny \yng(2)}$&1&2&0&$2r +\frac{2}{2n+1}$  \\
$\bar{B}^{(2,0)}:= \tilde{Q}^2$&1&1&${\tiny \yng(1,1)}$&0&2&$2 \bar{r}$  \\
$\bar{B}^{(1,1)}:= \tilde{Q} X \tilde{Q}$&1&1&${\tiny \yng(2)}$&0&2&$2 \bar{r} +\frac{2}{2n+1}$  \\  \hline
$Y_{SU(2)}$&1&1&1&$-F$&$-\bar{F}$&$F+\bar{F}-\frac{4}{2n+1}-Fr -\bar{F}r $  \\
$Y_{SU(2),X}$&1&1&1&$-F$&$-\bar{F}$&$F+\bar{F}-\frac{2}{2n+1}-Fr -\bar{F}r $  \\ \hline
  \end{tabular}}
  \end{center}\label{SU(2)KSelectric}
\end{table}

The magnetic description is given by the 3d $\mathcal{N}=2$ $SU(2nF-2)$ gauge theory with an adjoint $Y$, $F$ fundamental matters, $\bar{F}$ anti-fundamental matters and meson singlets $M_j~(j=0,1,\cdots,2n-1)$.
The theory includes a tree-level superpotential
\begin{align}
W_{mag}= \mathrm{tr} \, Y^{2n+1} + \sum_{j} M_j \tilde{q} Y^{2n-1-j} q.
\end{align}
For each decomposition of $F+\bar{F}$, there is a ``chiral'' dual description of this type. Therefore, the $SU(2)$ theory exhibits ``Seiberg N-ality'' \cite{Nii:2018bgf}. The matching of the generalized baryons is straightforward:
\begin{align}
B^{(2,0)} &\sim q^F (Yq)^F \cdots (Y^{2n-2}q)^{F} (Y^{2n-1}q)^{F-2} \\
B^{(1,1)} &\sim q^F (Yq)^F \cdots (Y^{2n-3}q)^{F} (Y^{2n-2}q)^{F-1} (Y^{2n-1}q)^{F-1}
\end{align}

Let us enumerate the Coulomb branch operators on the magnetic side. The magnetic Coulomb branch $\tilde{Y}_{a_m}^{bare}$ is possible only for $n \bar{F}-1 \le a_m \le n \bar{F}$ due to the constraints \eqref{acon1} and \eqref{acon2}. 
For $a_m =n \bar{F}$, the dressed operators are defined as
\begin{align}
Y_{a= n\bar{F}}^{dressed} &:=Y_{a=n \bar{F}}^{bare} \left(\tilde{q}^{\bar{F}} (Y \tilde{q})^{\bar{F}} \cdots (Y^{2n-1} \tilde{q})^{\bar{F}}  \right)^{\frac{F-\bar{F}}{2}} \\
 Y_{a= n\bar{F},Y}^{dressed} & :=Y_{a=n \bar{F}}^{bare}Y \left(\tilde{q}^{\bar{F}} (Y \tilde{q})^{\bar{F}} \cdots (Y^{2n-1} \tilde{q})^{\bar{F}}  \right)^{\frac{F-\bar{F}}{2}},
\end{align}
where $Y_{a= n\bar{F},Y}^{dressed}$ is dressed by the generalized anti-baryon and the $U(1)$ adjoint field. The non-abelian flavor indices of these operators are completely anti-symmetrized and these are flavor singlets. These are identified with the electric Coulomb branch operators $Y_{SU(2)}$ and $Y_{SU(2),X}$, respectively. 
For $a_m= n \bar{F}-1$, it is possible to make a more minimal Coulomb branch $\left( Y_{a=n\bar{F}-1}^{bare} \right)^{\frac{2}{F -\bar{F}}}$ gauge invariant by combining it with the generalized anti-baryon operators
\begin{align}
Y_{a=n \bar{F}-1}^{dressed,1} &:=\left( Y_{a=n\bar{F}-1}^{bare} \right)^{\frac{2}{F -\bar{F}}} \tilde{q}^{\bar{F}} \cdots (Y^{2n-2}\tilde{q})^{\bar{F}} (Y^{2n-1} \tilde{q})^{\bar{F}-2}  \\
{Y}_{a=n \bar{F}-1}^{dressed,2} &:=\left( Y_{a=n\bar{F}-1}^{bare} \right)^{\frac{2}{F -\bar{F}}} \tilde{q}^{\bar{F}} \cdots (Y^{2n-3}\tilde{q})^{\bar{F}} (Y^{2n-2} \tilde{q})^{\bar{F}-1}(Y^{2n-1}\tilde{q})^{\bar{F}-1}.
\end{align}
These are identified with the generalized anti-baryons $\bar{B}^{(2,0)}$ and $\bar{B}^{(1,1)}$, respectively. The quantum numbers of these operators are listed in Table \ref{SU(2)KSmagnetic}.

\begin{table}[H]\caption{$SU(2nF -2)$ magnetic dual description} 
\begin{center}
\scalebox{0.65}{
  \begin{tabular}{|c||c||c|c|c|c|c| } \hline
  &\small $SU(2nF-2)$&\small  $SU(F)$&\small  $SU(\bar{F})$& \small $U(1)$& \small  $U(1)$& \small  $U(1)_R$  \\ \hline
 $Y$ &$\mathbf{adj.}$&1&1&0&0&$\frac{2}{2n+1}$ \\
 $q$ & ${\tiny \yng(1)}$ &${\tiny \overline{\yng(1)}}$&1&$\frac{1}{nF-1}$&0&\footnotesize $\frac{r}{nF-1}-\frac{2n-1}{2n+1}\frac{nF-2}{nF-1}$ \\
$\tilde{q}$  &${\tiny \overline{\yng(1)}}$&1&${\tiny \overline{\yng(1)}}$&$\frac{-nF}{nF-1}$&$-1$& \footnotesize $\frac{-nFr}{nF-1}-\bar{r}+2-\frac{2n-1}{2n+1}\frac{nF}{nF-1}$ \\  
$M_{j}:=QX^j \tilde{Q}$&1&${\tiny \yng(1)}$&${\tiny \yng(1)}$&1&1&$r+\bar{r} +\frac{2j}{2n+1}$  \\  \hline
$T_2: \sim \mathrm{tr} \,Y ^2$&1&1&1&0&0&$\frac{4}{2n+1}$  \\ 
\scriptsize $B^{(2,0)} \sim q^F (Yq)^F \cdots (Y^{2n-2}q)^{F} (Y^{2n-1}q)^{F-2}$&1&${\tiny \yng(1,1)}$&1&2&0&$2r$  \\
\scriptsize $B^{(1,1)} \sim q^F (Yq)^F \cdots (Y^{2n-3}q)^{F} (Y^{2n-2}q)^{F-1} (Y^{2n-1}q)^{F-1}$&1&${\tiny \yng(2)}$&1&2&0&$2r +\frac{2}{2n+1}$  \\ \hline
\scriptsize $Y_{SU(2)} \sim Y_{a= n\bar{F}}^{dressed} :=Y_{a=n \bar{F}}^{bare} \left(\tilde{q}^{\bar{F}} (Y \tilde{q})^{\bar{F}} \cdots (Y^{2n-1} \tilde{q})^{\bar{F}}  \right)^{\frac{F-\bar{F}}{2}}$&1&1&1&$-F$&  $-\bar{F}$&\footnotesize$F+\bar{F}-\frac{4}{2n+1}-Fr -\bar{F}r $ \\
\scriptsize $Y_{SU(2),Y} \sim Y_{a= n\bar{F},Y}^{dressed} :=Y_{a=n \bar{F}}^{bare}Y \left(\tilde{q}^{\bar{F}} (Y \tilde{q})^{\bar{F}} \cdots (Y^{2n-1} \tilde{q})^{\bar{F}}  \right)^{\frac{F-\bar{F}}{2}}$&1&1&1&$-F$& $-\bar{F}$& \footnotesize $F+\bar{F}-\frac{4}{2n+1}-Fr -\bar{F}r $ \\
\scriptsize $\bar{B}^{(2,0)} \sim Y_{a=n \bar{F}-1}^{dressed,1}:=\left( Y_{a=n\bar{F}-1}^{bare} \right)^{\frac{2}{F -\bar{F}}} \tilde{q}^{\bar{F}} \cdots (Y^{2n-2}\tilde{q})^{\bar{F}} (Y^{2n-1} \tilde{q})^{\bar{F}-2}$&1&1&${\tiny \yng(1,1)}$&0&$2$&$2\bar{r}$ \\
\scriptsize $\bar{B}^{(1,1)} \sim {Y}_{a=n \bar{F}-1}^{dressed,2}:=\left( Y_{a=n\bar{F}-1}^{bare} \right)^{\frac{2}{F -\bar{F}}} \tilde{q}^{\bar{F}} \cdots (Y^{2n-3}\tilde{q})^{\bar{F}} (Y^{2n-2} \tilde{q})^{\bar{F}-1}(Y^{2n-1} \tilde{q})^{\bar{F}-1}$&1&1&${\tiny \yng(2)}$&0&2&$2\bar{r}+\frac{2}{2n+1}$ \\ \hline
  \end{tabular}}
  \end{center}\label{SU(2)KSmagnetic}
\end{table}

As a further check of the duality above, we investigate the case with $(F,\bar{F},2n)=(3,1,2)$ in detail. The electric side becomes the 3d $\mathcal{N}=2$ $SU(2)$ gauge theory with four fundamentals and an adjoint matter. The electric superpotential is again vanishing. The four fundamental matters are decomposed into three fundamental and one anti-fundamental matters. The non-vanishing moduli operators are listed in Table \ref{SU(2)31k2electric}. We emphasize that the dressed Coulomb branch \eqref{SU2u1dressed} is not truncated quantum-mechanically. This will be checked by the SCI calculation below.

\begin{table}[H]\caption{3d $\mathcal{N}=2$ $SU(2)$ with $\mathbf{adj.} + 3\, {\tiny \protect\yng(1)}+  \,{\tiny \overline{\protect\yng(1)}},~k=2$} 
\begin{center}
\scalebox{1}{
  \begin{tabular}{|c||c||c|c|c|c| } \hline
  &$SU(2)$&$SU(3)$&$U(1)$&$U(1)$&$U(1)_R$  \\ \hline
 $X$ &$\mathbf{adj.}$&1&0&0&$\frac{2}{3}$ \\
 $Q$ & ${\tiny \yng(1)}$ &${\tiny \yng(1)}$&1&0&$r$ \\
$\tilde{Q}$  &${\tiny \overline{\yng(1)}}={\tiny \yng(1)}$&1&$0$&1&$\bar{r}$ \\  \hline
$M_{j=0,1}:=QX^j \tilde{Q}$&1&${\tiny \yng(1)}$&1&1&$r+\bar{r} +\frac{2j}{3}$  \\
$T_{2}:=\mathrm{tr} \,X ^2$&1&1&0&0&$\frac{4}{3}$  \\ 
$B^{(2,0)}:=Q^2$&1&${\tiny \overline{\yng(1)}}$ &2&0&$2r$ \\
$B^{(1,1)}:=QXQ$&1&${\tiny \yng(2)}$ &2&0&$2r+\frac{2}{3}$\\
$\bar{B}^{(1,1)}:=\tilde{Q}X\tilde{Q}$&1&1&0&2&$2\bar{r}+\frac{2}{3}$ \\  \hline
$Y^{bare}_{SU(2)}$&1&1&$-3$&$-1$&$\frac{8}{3}-3r-\bar{r}$  \\
$Y_{SU(2),X}:=Y^{bare}_{SU(2)} X$&1&1&$-3$&$-1$&$\frac{10}{3}-3r-\bar{r}$  \\  \hline
  \end{tabular}}
  \end{center}\label{SU(2)31k2electric}
\end{table}

The magnetic description (see Table \ref{SU(2)31k2magnetic}) is the 3d $\mathcal{N}=2$ $SU(4)$ gauge theory with three fundamentals, an anti-fundamental, an adjoint matter and two gauge singlets $M_{j}\,(j=0,1)$. 
There are two types of Coulomb branches possible. The bare Coulomb branch $Y^{bare}_{SU(2)}$ corresponds to the gauge symmetry breaking $SU(4) \rightarrow SU(2) \times U(1)_1 \times U(1)_2$. Therefore, the dressed operators are defined as
\begin{align}
\tilde{Y}_{SU(2)}^{dressed} &:=\tilde{Y}^{bare}_{SU(2)} \tilde{q}(Y \tilde{q})   \\
 \tilde{Y}_{SU(2),Y}^{dressed} &:=\tilde{Y}^{bare}_{SU(2)} Y' \tilde{q}(Y \tilde{q}),
\end{align}
where $ \tilde{Y}_{SU(2),Y}^{dressed}$ is dressed by the $U(1)$ adjoint field $Y'$. One might consider that, since the low-energy gauge group contains the two $U(1)$ subgroups, there could be two $\tilde{Y}_{SU(2),Y}^{dressed}$ operators. However, from the index computation, we can see that the dressed fermionic operator
\begin{align}
\tilde{Y}^{bare}_{SU(2)} \tilde{q}(\psi_Y \tilde{q}) 
\end{align}
exactly cancels the additional $\tilde{Y}_{SU(2),Y}^{dressed}$ operator. This is consistent with our duality proposal since $\tilde{Y}_{SU(2),Y}^{dressed}$ is identified with $\tilde{Y}_{SU(2),X}$ and the electric side only allows a single $Y_{SU(2),X}$ operator. This is an example of the quantum-mechanical truncation of the chiral ring elements. The second Coulomb branch $\tilde{Y}_{SU(2) \times SU(2)}$ corresponds to the breaking $SU(4) \rightarrow SU(2) \times SU(2) \times U(1)$. The bare operator $\tilde{Y}_{SU(2) \times SU(2)}$ is gauge invariant and can be used for the moduli coordinate. In this case, although it is again possible to define the operator dressed by the $U(1)$ adjoint scalar, we observed that the lowest contribution of the index with a GNO charge $(1,1,0,-2)$ cancels this operator. This is again consistent with our duality proposal since $\tilde{Y}_{SU(2) \times SU(2)}$ is identified with $\bar{B}^{(1,1)}$ and $\tilde{Y}_{SU(2) \times SU(2)} Y_{U(1)}$ has no counterpart on the electric side. The similar cancelation was observed in the 3d $\mathcal{N}=2$ ``chiral'' $U(N)$ gauge theory \cite{Aharony:2015pla}.

\begin{table}[H]\caption{$SU(4)$ magnetic dual of Table \ref{SU(2)31k2electric}} 
\begin{center}
\scalebox{0.95}{
  \begin{tabular}{|c||c||c|c|c|c| } \hline
  &$SU(4)$&$SU(3)$&$U(1)$&$U(1)$&$U(1)_R$  \\ \hline
 $Y$ &$\mathbf{adj.}$&1&0&0&$\frac{2}{3}$ \\
 $q$ & ${\tiny \yng(1)}$ &${\tiny \overline{\yng(1)}}$&$\frac{1}{2}$&0&$\frac{1}{2}r$ \\
$\tilde{q}$  &${\tiny \overline{\yng(1)}}$&1&$-\frac{3}{2}$&$-1$&$\frac{3}{2}-\frac{3}{2}r-\bar{r}$ \\ 
$M_{j=0,1}$&1&${\tiny \yng(1)}$&1&1&$r+\bar{r} +\frac{2j}{3}$  \\ \hline
$T_{2} \sim \mathrm{tr} \,Y^2$&1&1&0&0&$\frac{4}{3}$  \\ 
$B^{(2,0)} \sim q^3(Yq)$&1&${\tiny \overline{\yng(1)}}$ &2&0&$2r$ \\
$B^{(1,1)} \sim q^2(Yq)^2$&1&${\tiny \yng(2)}$ &2&0&$2r+\frac{2}{3}$\\ \hline
$\tilde{Y}^{bare}_{SU(2)}$&$U(1)_2:-2$&1&$0$&$1$&$\bar{r} -1 $  \\
$Y^{bare}_{SU(2)} \sim \tilde{Y}_{SU(2)}^{dressed}:=\tilde{Y}^{bare}_{SU(2)} \tilde{q}(Y \tilde{q})$&1&1&$-3$&$-1$&$\frac{8}{3}-3r-\bar{r}$  \\ 
$Y_{SU(2),X} \sim \tilde{Y}_{SU(2),Y}^{dressed}:=\tilde{Y}^{bare}_{SU(2)} Y' \tilde{q}(Y \tilde{q})$&1&1&$-3$&$-1$& $\frac{10}{3}-3r-\bar{r}$   \\  \hline
$\bar{B}^{(1,1)} \sim \tilde{Y}_{SU(2) \times SU(2)}$&1&1&$0$&2&$2\bar{r}+\frac{2}{3}$  \\ \hline
  \end{tabular}}
  \end{center}\label{SU(2)31k2magnetic}
\end{table}

The electric and magnetic indices are computed by employing the localization method \cite{Bhattacharya:2008bja, Kim:2009wb, Imamura:2011su, Kapustin:2011jm} and we observed a nice agreement. Due to the limitation of the machine power, we computed the magnetic indices up to $O(x^2)$. The index takes the following form

\footnotesize
\begin{align}
\begin{autobreak}
I=1
+\frac{x^{2/3}}{t^3 u}
+3 x \left(t^2+t u\right)
+x^{4/3} \left(\frac{t^3 u+1}{t^6 u^2}+1\right)
+x^{5/3} \left(6 t^2+3 t u+u^2\right)
+x^2 \left(6 t^4+8 t^3 u+6 t^2 u^2+\frac{t^3 u+1}{t^9 u^3}-\frac{3 t}{u}-\frac{3 u}{t}-11\right)
+x^{7/3} \left(-3 t^2-u^2\right)
+x^{8/3} \left(15 t^4+18 t^3 u+9 t^2 u^2+\frac{t^3 u+1}{t^{12} u^4}+3 t u^3-\frac{3 t}{u}-\frac{3 u}{t}-9\right)
+x^3 \left(10 t^6+15 t^5 u+15 t^4 u^2+10 t^3 u^3-\frac{8 t^3}{u}-27 t^2+\frac{6}{t^2}-27 t u+\frac{3}{t u}-8 u^2+\frac{1}{u^2}\right)
+x^{10/3} \left(6 t^4+3 t^2 u^2+\frac{t^{12} u^4+t^3 u+1}{t^{15} u^5}+u^4+1\right)
+x^{11/3} \left(27 t^6+39 t^5 u+36 t^4 u^2+18 t^3 u^3-\frac{18 t^3}{u}+\frac{u}{t^3}+6 t^2 u^4-57 t^2+\frac{9}{t^2}-\frac{3 u^3}{t}-42 t u+\frac{12}{t u}-18 u^2\right)
+x^4 \biggl(15 t^8+24 t^7 u+27 t^6 u^2+24 t^5 u^3-\frac{15 t^5}{u}+15 t^4 u^4-63 t^4-64 t^3 u-54 t^2 u^2+\frac{3 t^2}{u^2}+\frac{3 u^2}{t^2}
\qquad +\frac{t^{12} u^4+t^3 u+1}{t^{18} u^6}-18 t u^3+\frac{27 t}{u}+\frac{27 u}{t}-u^4+42\biggr)
+\cdots,
\end{autobreak}
\end{align}
\normalsize
where the r-charges of the (anti-)fundamental matters are fixed to $r=\bar{r}=\frac{1}{2}$ for simplicity. $t$ and $u$ are the fugacities for the two $U(1)$ global symmetries. The second term $\frac{x^{2/3}}{t^3 u}$ corresponds to the Coulomb branch operator $Y_{SU(2)}$. The third term $3 x \left(t^2+t u\right)$ consists of $B^{(2,0)}$ and $M_0$. The fourth term $x^{4/3} \left(\frac{t^3 u+1}{t^6 u^2}+1\right)$ is interpreted as the contribution of $Y_{SU(2),X} +Y_{SU(2)}^2 +T_2$. The fifth term $x^{5/3} \left(6 t^2+3 t u+u^2\right)$ is identified with $B^{(1,1)} +M_1+\bar{B}^{(1,1)}$. Notice that, at $O(x^{5/3})$, we cannot consider the operators like $Y_{SU(2)} (M_0 +B^{(2,0)})$ since all the components of the (anti-)fundamental matters are massive along the Coulomb branch. The higher order terms are regarded as the symmetric products of these bosonic operators and the fermion contributions.

\subsection{$SU(3)$ self-dual example}
Next, we investigate the 3d $\mathcal{N}=2$ $SU(3)$ gauge theory with an adjoint matter $X$, three fundamental matters $Q$ and a single anti-fundamental matter $\tilde{Q}$. The theory includes a tree-level superpotential
\begin{align}
W_{ele}= \mathrm{tr} \, X^3.
\end{align}
The Higgs branch is completely the same as the 4d one except for the absence of the anti-baryonic operators. The Higgs branch operators are defined by
\begin{gather*}
M_{j}:=QX^j \tilde{Q}~~(j=0,1),~~~T_2:=\mathrm{tr} \,X ^2,~~~B^{(3,0)}:=Q^3,~~~B^{(2,1)}:=Q^2(XQ).
\end{gather*}
When the bare Coulomb branch $Y^{bare}$ obtains a non-zero expectation value, the gauge group is broken to
\begin{align}
SU(3) & \rightarrow U(1)_1  \times U(1)_2 \\
\mathbf{3} & \rightarrow  \mathbf{1}_{1,1} +\mathbf{1}_{0,-2} +\mathbf{1}_{-1,1} \\
\overline{\mathbf{3}} & \rightarrow  \mathbf{1}_{-1,-1} +\mathbf{1}_{0,2} +\mathbf{1}_{1,-1} \\
\mathbf{8} & \rightarrow  \mathbf{1}_{0,0} +\mathbf{1'}_{0,0} +\mathbf{1}_{\pm 2,0} +\mathbf{1}_{\pm 1,\pm 3}.
\end{align}
Along the Coulomb branch, some components of the (anti-)fundamental matters are massive and integrated out, which results in the mixed Chern-Simons term between the two $U(1)$ gauge symmetries. Therefore, the bare Coulomb branch $Y^{bare}$ has the non-zero charge under the $U(1)_2$ symmetry. Thus, we have to define the following dressed (gauge invariant) operators
\begin{align}
Y^{dressed}&:=Y^{bare} \mathbf{1}_{0,2} \sim Y^{bare} \tilde{Q} \\
Y^{dressed}_{X}&:=Y^{bare}  \mathbf{1}_{0,0}  \mathbf{1}_{0,2}  \sim Y^{bare} X\tilde{Q} \\
Y^{dressed}_{X'}&:= Y^{bare}  \mathbf{1'}_{0,0}  \mathbf{1}_{0,2} \sim Y^{bare} X\tilde{Q}.
\end{align}
Notice that there are two types of operators dressed by $X$ since, along the Coulomb branch, the adjoint matter reduces to the two massless gauge singlets, $\mathbf{1}_{0,0}$ and $\mathbf{1'}_{0,0}$. Table \ref{SU(3)SDelectric} summarizes the quantum numbers of the elementary fields and the moduli operators in the electric description.

\begin{table}[H]\caption{3d $\mathcal{N}=2$ $SU(3)$ with $\mathbf{adj.} + 3\, {\tiny \protect\yng(1)}+  \,{\tiny \overline{\protect\yng(1)}},~W_{ele}= \mathrm{tr} \, X^3$} 
\begin{center}
\scalebox{1}{
  \begin{tabular}{|c||c||c|c|c|c| } \hline
  &$SU(3)$&$SU(3)$&$U(1)$&$U(1)$&$U(1)_R$  \\ \hline
 $X$ &$\mathbf{adj.}$&1&0&0&$\frac{2}{3}$ \\
 $Q$ & ${\tiny \yng(1)}$ &${\tiny \yng(1)}$&1&0&$r$ \\
$\tilde{Q}$  &${\tiny \overline{\yng(1)}}$&1&$0$&1&$\bar{r}$ \\  \hline
$M_{j=0,1}:=QX^j \tilde{Q}$&1&${\tiny \yng(1)}$&1&1&$r+\bar{r} +\frac{2j}{3}$  \\
$T_2:=\mathrm{tr} \,X ^2$&1&1&0&0&$\frac{4}{3}$  \\ 
$B^{(3,0)}:=Q^3$&1&1&3&0&$3r$ \\
$B^{(2,1)}:=Q^2(XQ)$&1& ${\tiny \yng(2,1)}$&3&0&$3r+\frac{2}{3}$  \\ \hline
$Y^{bare}$&$U(1)_2:-2$&1&$-3$&$-1$&$\frac{4}{3}-3r-\bar{r}$  \\
$Y^{dressed}:=Y^{bare} \tilde{Q}$&1&1&$-3$&$0$&$\frac{4}{3}-3r$  \\
$Y^{dressed}_{X}:=Y^{bare} X\tilde{Q}$&1&1&$-3$&$0$&$2-3r$ \\
$Y^{dressed}_{X'}:=Y^{bare} X\tilde{Q}$&1&1&$-3$&$0$&$2-3r$ \\ \hline
  \end{tabular}}
  \end{center}\label{SU(3)SDelectric}
\end{table}

Let us consider the ``chiral'' Kutasov-Schwimmer dual description of the above theory. The dual is given by the 3d $\mathcal{N}=2$ $SU(3)$ gauge theory with an adjoint matter $Y$, three fundamental matters $q$, an anti-fundamental matter $\tilde{q}$ and two meson singlets $M_{j=0,1}$. This is an example of the self-duality \cite{Csaki:1997cu, Karch:1997jp, Nii:2018uck}. The dual theory includes a tree-level superpotential
\begin{align}
W_{mag}= \mathrm{tr} \, Y^3 +M_0 \tilde{q}Yq +M_1  \tilde{q}q,
\end{align}
which truncates the chiral ring of $Y$ and lifts the dual mesons. The matter content and their quantum numbers are summarized in Table \ref{SU(3)SDmagnetic}. The $U(1)$ charge assignment is fixed by the above superpotential and by requiring the baryon matching. The analysis of the Higgs and Coulomb branches is the same as the electric one. From Table \ref{SU(3)SDelectric} and Table \ref{SU(3)SDmagnetic}, we can easily see the matching of the gauge invariant operators. 

\begin{table}[H]\caption{$SU(3)$ self-dual description} 
\begin{center}
\scalebox{1}{
  \begin{tabular}{|c||c||c|c|c|c| } \hline
  &$SU(3)$&$SU(3)$&$U(1)$&$U(1)$&$U(1)_R$  \\ \hline
 $Y$ &$\mathbf{adj.}$&1&0&0&$\frac{2}{3}$ \\
 $q$ & ${\tiny \yng(1)}$ &${\tiny \overline{\yng(1)}}$&1&0&$r$ \\
$\tilde{q}$  &${\tiny \overline{\yng(1)}}$&1&$-2$&$-1$&$\frac{4}{3}-2r-\bar{r}$ \\ 
$M_{j=0,1}$&1&${\tiny \yng(1)}$&1&1&$r+\bar{r} +\frac{2j}{3}$  \\ \hline
$T_2 \sim \mathrm{tr} \,Y ^2$&1&1&0&0&$\frac{4}{3}$  \\ 
$B^{(3,0)} \sim q^3$&1&1&3&0&$3r$ \\
$B^{(2,1)}  \sim q^2(Yq)$&1& ${\tiny \yng(2,1)}$&3&0&$3r+\frac{2}{3}$  \\ \hline
$\tilde{Y}^{bare}$&$U(1)_2:-2$&1&$-1$&$1$&$-r+\bar{r}$  \\
$Y^{dressed} \sim \tilde{Y}^{dressed}:=\tilde{Y}^{bare} \tilde{q}$&1&1&$-3$&$0$&$\frac{4}{3}-3r$  \\
$Y^{dressed}_{X} \sim \tilde{Y}^{dressed}_{Y}:= \tilde{Y}^{bare} Y\tilde{q}$&1&1&$-3$&$0$&$2-3r$ \\
$Y^{dressed}_{X'} \sim \tilde{Y}^{dressed}_{Y'}:=\tilde{Y}^{bare} Y\tilde{q}$&1&1&$-3$&$0$&$2-3r$ \\ \hline
  \end{tabular}}
  \end{center}\label{SU(3)SDmagnetic}
\end{table}

Finally, we will test the $SU(3)$ self-duality by computing the superconformal indices \cite{Bhattacharya:2008bja, Kim:2009wb, Imamura:2011su, Kapustin:2011jm} of the duality pair discussed above. We computed the indices up to $O(x^3)$ and found a perfect agreement. The index becomes
\scriptsize
\begin{align}
\begin{autobreak}
I=1+
3 t u \sqrt{x}
+\frac{x^{7/12}}{t^3}
+t^3 x^{3/4}
+6 t^2 u^2 x
+\frac{3 u x^{13/12}}{t^2}
+x^{7/6} \left(\frac{1}{t^6}+3 t u\right)
+x^{5/4} \left(3 t^4 u+\frac{2}{t^3}\right)
+x^{4/3}
+8 t^3 x^{17/12}
+x^{3/2} \left(t^6+10 t^3 u^3\right)
+\frac{6 u^2 x^{19/12}}{t}
+x^{5/3} \left(\frac{3 u}{t^5}+9 t^2 u^2\right)
+x^{7/4} \left(\frac{1}{t^9}+6 t^5 u^2+\frac{6 u}{t^2}\right)
+x^{11/6} \left(\frac{2}{t^6}+3 t u\right)+x^{23/12} \left(24 t^4 u+\frac{1}{t^3}\right)
+x^2 \left(3 t^7 u+15 t^4 u^4-10\right)
+x^{25/12} \left(-\frac{3}{t^4 u}+t^3+10 u^3\right)
+x^{13/6} \left(8 t^6+\frac{6 u^2}{t^4}+18 t^3 u^3\right)
+x^{9/4} \left(t^9+\frac{3 u}{t^8}+10 t^6 u^3-\frac{3 t^2}{u}+\frac{12 u^2}{t}\right)
+x^{7/3} \left(\frac{1}{t^{12}}+\frac{6 u}{t^5}+12 t^2 u^2\right)
+x^{29/12} \left(\frac{2}{t^9}+48 t^5 u^2+\frac{3 u}{t^2}\right)
+x^{5/2} \left(6 t^8 u^2+\frac{1}{t^6}+21 t^5 u^5-27 t u\right)
+x^{31/12} \left(18 t^4 u-\frac{9}{t^3}+15 t u^4\right)
+x^{8/3} \left(24 t^7 u-\frac{3}{t^7 u}+30 t^4 u^4+\frac{10 u^3}{t^3}-10\right)
+x^{11/4} \left(3 t^{10} u+15 t^7 u^4+\frac{6 u^2}{t^7}-\frac{6}{t^4 u}-17 t^3+20 u^3\right)
+x^{17/6} \left(\frac{3 u}{t^{11}}+\frac{12 u^2}{t^4}+28 \left(t^6+t^3 u^3\right)-\frac{3}{t u}\right)
+x^{35/12} \left(\frac{1}{t^{15}}+8 t^9+\frac{6 u}{t^8}+80 t^6 u^3-\frac{9 t^2}{u}+\frac{3 u^2}{t}\right)
+x^3 \left(t^{12}+\frac{2}{t^{12}}+10 t^9 u^3+28 t^6 u^6-\frac{3 t^5}{u}+\frac{3 u}{t^5}-51 t^2 u^2\right)+\cdots,
\end{autobreak}
\end{align}
\normalsize
where $t$ and $u$ are the fugacities for the global $U(1)$ symmetries. The r-charges are set to $r=\bar{r}=\frac{1}{4}$ for simplicity. The second term $3 t u \sqrt{x}$ corresponds to the meson $M_0$. The third term $\frac{x^{7/12}}{t^3}$ is the dressed Coulomb branch $Y^{dressed}$. The fourth term $t^3 x^{3/4}$ is identified with $B^{(3,0)}$. The fifth term $6 t^2 u^2 x$ is a symmetric product of $M_0$. The seventh term $x^{7/6} \left(\frac{1}{t^6}+3 t u\right)$ is the sum of $(Y^{dressed})^2$ and $M_1$. The two Coulomb branch operators, $Y^{dressed}_{X}$ and $Y^{dressed}_{X'} $ are expressed as $x^{5/4} \frac{2}{t^3}$, which is consistent with our analysis of the dressed Coulomb branch. The ninth and tenth terms $x^{4/3}+8 t^3 x^{17/12}$ correspond to $T_2 + B^{(2,1)}$.

\subsection{$SU(3)$ with $(F,\bar{F},k)=(4,2,2)$}
As a final example, we study another example of the $SU(3)$ duality. The electric theory is the 3d $\mathcal{N}=2$ $SU(3)$ gauge theory with an adjoint matter $X$, four fundamental matters $Q$ and two anti-fundamental matters $\tilde{Q}$. The theory includes a tree-level superpotential
\begin{align}
W_{ele} = \mathrm{tr} \, X^3.
\end{align}
The Higgs and Coulomb branch is the same as the previous one except for the presence of the generalized anti-baryon operator
\begin{align}
\bar{B}^{(2,1)}:= \tilde{Q}^2(X\tilde{Q}).
\end{align}
The bare Coulomb branch induces the gauge symmetry breaking $SU(3) \rightarrow U(1)_1 \times U(1)_2$  and obtains a non-zero $U(1)_2$ charge via the mixed Chern-Simons term between $U(1)_1$ and $U(1)_2$. As a result, the gauge invariant operators are defined by
\begin{align}
Y^{dressed}_{\tilde{Q}} &:=Y^{bare} \tilde{Q},\\
Y^{dressed}_{\tilde{Q}X} &:=Y^{bare} \tilde{Q}X \\
Y^{dressed}_{\tilde{Q}X'} &:=Y^{bare} \tilde{Q}X',
\end{align}
where we can introduce two operators dressed by the adjoint scalar since the low-energy theory has the two $U(1)$ subgroups along the Coulomb branch. Table \ref{SU(3)42k2electric} summarizes the quantum numbers of the elementary fields and the gauge invariant operators.

\begin{table}[H]\caption{3d $\mathcal{N}=2$ $SU(3)$ with $\mathbf{adj.} + 4\, {\tiny \protect\yng(1)}+ 2 \,{\tiny \overline{\protect\yng(1)}},~W_{ele} = \mathrm{tr} \, X^3$} 
\begin{center}
\scalebox{0.9}{
  \begin{tabular}{|c||c||c|c|c|c|c| } \hline
  &$SU(3)$&$SU(4)$&$SU(2)$&$U(1)$&$U(1)$&$U(1)_R$  \\ \hline
 $X$ &$\mathbf{adj.}$&1&1&0&0&$\frac{2}{3}$ \\
 $Q$ & ${\tiny \yng(1)}$ &${\tiny \yng(1)}$&1&1&0&$r$ \\
$\tilde{Q}$  &${\tiny \overline{\yng(1)}}$&1&${\tiny \yng(1)}$&$0$&1&$\bar{r}$ \\  \hline
$M_{j=0,1}:=QX^j \tilde{Q}$&1&${\tiny \yng(1)}$&${\tiny \yng(1)}$&1&1&$r+\bar{r} +\frac{2j}{3}$  \\
$T_2:=\mathrm{tr} \,X ^2$&1&1&1&0&0&$\frac{4}{3}$  \\ 
$B^{(3,0)} :=Q^3$&1&${\tiny \overline{\yng(1)}}$&1&3&0&$3r$  \\
$B^{(2,1)} :=Q^2(XQ)$&1&${\tiny \yng(2,1)}$&1&3&0&$3r +\frac{2}{3}$  \\
$\bar{B}^{(2,1)}:= \tilde{Q}^2(X\tilde{Q})$&1&1&${\tiny \yng(1)}$&0&3&$3 \bar{r}+\frac{2}{3}$  \\ \hline
$Y^{bare}$&$U(1)_2$: $-2$&1&1&$-4$&$-2$&$\frac{10}{3}-4r -2\bar{r} $  \\
$Y^{dressed}_{\tilde{Q}}:=Y^{bare} \tilde{Q}$&1&1&${\tiny \yng(1)}$&$-4$&$-1$&$\frac{10}{3}-4r -\bar{r}  $  \\
$Y^{dressed}_{\tilde{Q}X}:=Y^{bare} \tilde{Q}X$&1&1&${\tiny \yng(1)}$&$-4$&$-1$&$4-4r -\bar{r}  $ \\
$Y^{dressed}_{\tilde{Q}X'}:=Y^{bare} \tilde{Q}X'$&1&1&${\tiny \yng(1)}$&$-4$&$-1$&$4-4r -\bar{r}  $ \\ \hline
  \end{tabular}}
  \end{center}\label{SU(3)42k2electric}
\end{table}

The magnetic description is given by the 3d $\mathcal{N}=2$ $SU(5)$ gauge theory with an adjoint matter $Y$, four fundamental matters $q$, two anti-fundamental matters $\tilde{q}$ and two meson singlets $M_{j},~(j=0,1)$. The dual theory includes the tree-level superpotential
\begin{align}
W_{mag} = \mathrm{tr} \, Y^3  + M_0 \tilde{q} Y q + M_1 \tilde{q}q,
\end{align}
which lifts the dual mesons and truncates the chiral ring elements constructed from $Y$. The matter fields and their quantum numbers are summarized in Table \ref{SU(3)42k2magnetic}. This example is very illustrative since the $SU(5)$ magnetic theory does not allow the anti-baryonic operators. The missing operator $\bar{B}^{(2,1)}$ will come from the dual Coulomb branch. 

The first Coulomb branch $\tilde{Y}_{SU(3)}^{bare}$ corresponds to the gauge symmetry breaking
\begin{align}
SU(5) &\rightarrow SU(3) \times U(1)_1 \times U(1)_2\\
\mathbf{5} & \rightarrow  \mathbf{3}_{0,-2}+\mathbf{1}_{1,3}+\mathbf{1}_{-1,3}      \\
\overline{\mathbf{5}} & \rightarrow   \overline{\mathbf{3}}_{0,2}+\mathbf{1}_{-1,-3}+\mathbf{1}_{1,-3}  \\
\mathbf{24} & \rightarrow \mathbf{8}_{0,0}+\mathbf{1}_{0,0}+\mathbf{1}_{0,0}+\mathbf{3}_{1,-5}+\mathbf{3}_{-1,-5}+\overline{\mathbf{3}}_{-1,5}+\overline{\mathbf{3}}_{1,5}+\mathbf{1}_{2,0}+\mathbf{1}_{-2,0}
\end{align}
Due to the mixed Chern-Simons term between the $U(1)_1$ and $U(1)_2$ gauge groups, the bare operator $\tilde{Y}_{SU(3)}^{bare}$ obtains a non-zero $U(1)_2$ charge. The dressed gauge invariant operators are defined by
\begin{align}
Y^{dressed}_{SU(3),\tilde{q}} &:=\tilde{Y}_{SU(3)}^{bare}\tilde{q}^2 (Y \tilde{q})  \\
Y^{dressed}_{SU(3),\tilde{q}} &:=\tilde{Y}_{SU(3)}^{bare}\tilde{q}^2 (Y \tilde{q}) Y  \\
 Y^{dressed}_{SU(3),\tilde{q}} &:=\tilde{Y}_{SU(3)}^{bare}\tilde{q}^2 (Y \tilde{q}) Y',
\end{align}
where the last two operators are dressed by the $U(1)_1$ and $U(1)_2$ adjoint scalars, respectively. These are identified with the electric Coulomb branch operators $Y^{dressed}_{\tilde{Q}}, Y^{dressed}_{\tilde{Q}X}$ and $Y^{dressed}_{\tilde{Q}X'}$.

When the second Coulomb branch $\tilde{Y}^{bare}_{SU(2)^2}$ obtains a non-zero expectation value, the gauge group and the elementary fields are decomposed into
\begin{align}
SU(5) &\rightarrow SU(2) \times SU(2) \times U(1)_1 \times U(1)_2 \\
\mathbf{5} & \rightarrow (\mathbf{2}, \mathbf{1})_{1,1}+(\mathbf{1},\mathbf{2})_{-1,1}+(\mathbf{1},\mathbf{1})_{0,-4} \\
\overline{\mathbf{5}} & \rightarrow (\mathbf{2}, \mathbf{1})_{-1,-1}+(\mathbf{1},\mathbf{2})_{1,-1}+(\mathbf{1},\mathbf{1})_{0,4} \\
\mathbf{24} & \rightarrow (\mathbf{3}, \mathbf{1})_{0,0}+(\mathbf{1},\mathbf{3})_{0,0}+2(\mathbf{1},\mathbf{1})_{0,0} \nonumber \\& \qquad+(\mathbf{2}, \mathbf{2})_{ \pm 2,0} +(\mathbf{2}, \mathbf{1})_{ 1, 5}+(\mathbf{2}, \mathbf{1})_{-1, -5}+(\mathbf{1}, \mathbf{2})_{- 1, 5}+(\mathbf{1}, \mathbf{2})_{1, -5}.
\end{align}
Since the mixed Chern-Simons term is generated along the RG flow of this breaking, the bare Coulomb branch has to be dressed by the massless component $(\mathbf{1},\mathbf{1})_{0,4} $ from the anti-fundamental matter 
\begin{align}
\tilde{Y}^{dressed}_{SU(2)^2}:=\tilde{Y}^{bare}_{SU(2)^2} \tilde{q}.
\end{align}
Although it is possible to classically construct the operator dressed by the $U(1)$ adjoint scalars, those contributions are canceled by the lowest order contribution of the index with a GNO charge $(1,1,0,0,-2)$. This observation is consistent with the analysis on the electric side because $\tilde{Y}^{dressed}_{SU(2)^2}$ is identified with the anti-baryon $\bar{B}^{(2,1)} $.

\begin{table}[H]\caption{$SU(5)$ magnetic dual description} 
\begin{center}
\scalebox{0.85}{
  \begin{tabular}{|c||c||c|c|c|c|c| } \hline
  &$SU(5)$&$SU(4)$&$SU(2)$&$U(1)$&$U(1)$&$U(1)_R$  \\ \hline
 $Y$ &$\mathbf{adj.}$&1&1&0&0&$\frac{2}{3}$ \\
 $q$ & ${\tiny \yng(1)}$ &${\tiny \overline{\yng(1)}}$&1&$\frac{3}{5}$&0&$-\frac{2}{15}+\frac{3}{5}r$ \\
$\tilde{q}$  &${\tiny \overline{\yng(1)}}$&1&${\tiny \yng(1)}$&$-\frac{8}{5}$&$-1$&$\frac{22}{15}-\frac{8}{5}r-\bar{r}$ \\ 
$M_j ~(j=0,1)$&1&${\tiny \yng(1)}$&${\tiny \yng(1)}$&1&1&$r+\bar{r} +\frac{2j}{3}$  \\ \hline
$T_2 \sim \mathrm{tr} \, Y^2$&1&1&1&0&0&$\frac{4}{3}$  \\ 
$B^{(3,0)} \sim q^4 (Yq)$&1&${\tiny \overline{\yng(1)}}$&1&3&0&$3r$  \\
$B^{(2,1)} \sim q^3(Yq)^2$&1&${\tiny \yng(2,1)}$&1&3&0&$3r +\frac{2}{3}$  \\  \hline
$\tilde{Y}_{SU(3)}^{bare}$&$U(1)_2$: $-6$&1&1&$\frac{4}{5}$&$2$&$\frac{-26}{15}+\frac{4}{5}r +2\bar{r} $  \\
$Y^{dressed}_{\tilde{Q}}\sim Y^{dressed}_{SU(3),\tilde{q}}:=\tilde{Y}_{SU(3)}^{bare}\tilde{q}^2 (Y \tilde{q})$&1&1&${\tiny \yng(1)}$&$-4$&$-1$&$\frac{10}{3}-4r -\bar{r}  $  \\
$Y^{dressed}_{\tilde{Q}X} \sim Y^{dressed}_{SU(3),\tilde{q}Y}:=\tilde{Y}_{SU(3)}^{bare}\tilde{q}^2 (Y \tilde{q}) Y$&1&1&${\tiny \yng(1)}$&$-4$&$-1$&$4-4r -\bar{r}  $ \\
$Y^{dressed}_{\tilde{Q}X'} \sim Y^{dressed}_{SU(3),\tilde{q}Y'}:=\tilde{Y}_{SU(3)}^{bare}\tilde{q}^2 (Y \tilde{q}) Y'$&1&1&${\tiny \yng(1)}$&$-4$&$-1$&$4-4r -\bar{r}  $ \\ \hline
$\tilde{Y}^{bare}_{SU(2)^2}$&$U(1)_2$: $-4$&1&1&$\frac{8}{5}$&$4$&$-\frac{12}{5}+\frac{8}{5} r +4\bar{r}$  \\
$\bar{B}^{(2,1)} \sim \tilde{Y}^{dressed}_{SU(2)^2}:=\tilde{Y}^{bare}_{SU(2)^2} \tilde{q}$&1&1&${\tiny \yng(1)}$&0&$3$&$\frac{2}{3}+3\bar{r}$  \\  \hline
  \end{tabular}}
  \end{center}\label{SU(3)42k2magnetic}
\end{table}

\section{Summary and Discussion}

In this paper, we proposed the ``chiral'' version of the 3d Kutasov-Schwimmer duality in the 3d $\mathcal{N}=2$ $SU(N)$ gauge theory with $F$ fundamentals, $\bar{F}$ anti-fundamentals and an adjoint matter including the three-level superpotential $W=\mathrm{tr} \, X^{k+1}$. By assuming $F>\bar{F}$, the dual description is given by the 3d $\mathcal{N}=2$ $SU(kF-N)$ gauge theory with $F$ fundamentals, $\bar{F}$ anti-fundamentals and an adjoint matter, including the three-level superpotential $W=\mathrm{tr} \, Y^{k+1} + \sum_j M_j \tilde{q} Y^{k-1-j}q$. We analyzed the structure of the (dressed) Coulomb branch operators of the moduli space of vacua and found the correct operator matching under the duality. In order to verify the validity of the duality, it was important to notice that some Coulomb branch operators are mapped to the generalized anti-baryonic operators in the magnetic theory.    
For the consistency check of the proposed duality, we considered several examples of the duality, which include $SU(2)$, $SU(3)$, $SU(4)$ and $SU(5)$ gauge groups. For some cases, we computed the superconformal indices and found the agreement of our duality. When the dual gauge group is vanishing, we found the s-confinement phase.

There are various future directions which we have to further investigate.
The 3d dualities were first developed in the ``vector-like'' theories. These dualities are related to each other and some dualities are derived from the 4d dualities \cite{Aharony:2013dha, Aharony:2013kma} via dimensional reduction and some deformations. It would be important to find the flow between the vector-like and chiral dualities discussed here. This can be achieved by introducing real masses to fundamental matters. However, the RG flow is not so simple on the magnetic side where we have to take a non-trivial point of the moduli space and flow to the low-energy limit. It is also important to find the flow from the duality studied here to the Chern-Simons duality \cite{Niarchos:2008jb, Niarchos:2009aa}.  

We mostly focused on the non-abelian Coulomb phase, where the moduli space has some singularities, and we did not exhaust all the possibilities of the confinement phases. Although we expect that the s-confinement phase appears when the rank of the dual gauge group becomes zero, we could not derive a general form of the confining superpotential. In these confining examples, the UV r-charges of the moduli coordinates are sometimes negative and hence we cannot rely on the perturbative calculation of the SCI around $x=0$. This implies that the IR r-symmetry would be mixed with some emergent symmetries. It would be nice if we gain a better understanding of the confinement (or SUSY breaking) phases in the ``chiral'' $SU(N)$ Kutasov-Schwimmer duality.

In this paper, we discussed the na\"ive matching of the gauge invariant operators under the duality. However, these operators are occasionally constrained or related to each other classically and quantum-mechanically. Generally, the classical constraints of the gauge invariant operators on the electric side can be regarded as the quantum relations on the magnetic side \cite{Kutasov:1995ss}. In the several examples, we observed the non-perturbative truncation of the Coulomb branch dressed by $U(1)$ adjoint scalars. However, we did not generally discuss the matching of the classical and quantum constraints under the duality. It is valuable to systematically find the non-perturbative constraints and find the matching of the quantum chiral ring structure under the duality transformation. 

It is curious to generalize our ``chiral'' duality to more general cases. For instance, one can consider the $SU(N)$ gauge theory with multiple adjoint matters and chiral (anti-)fundamentals. In 4d, the Kutasov-Schwimmer duality was generalized along this direction \cite{Brodie:1996vx, Brodie:1996xm} while in 3d the vector-like Kutasov-Schwimmer duality with two adjoint was recently discussed in \cite{Hwang:2018uyj}. It would be possible to extend our chiral duality along this direction.

\section*{Acknowledgments}
This work is supported by the Swiss National Science Foundation (SNF) under grant number PP00P2\_157571/1 and by ``The Mathematics of Physics'' (SwissMAP) under grant number NCCR 51NF40-141869.


\bibliographystyle{ieeetr}
\bibliography{chiralKSreferences}

\end{document}